\documentclass{aastex631}

\shorttitle{Morphology and kinematics of the gas in M51}
\shortauthors{Font et al.}
\graphicspath{{./}{figures/}}

\begin{document}

\title{Morphology and kinematics of the gas in M51: How interaction with NGC5195 has moulded the structure of its arms.}

\author{Joan Font}
\affiliation{Gemini Observatory/NSF's NOIRLab. Casilla 6035, La Serena. Chile \\}

\author{John E. Beckman}
\affiliation{Instituto de Astrof\'{i}sica de Canarias, 38205. La Laguna. Tenerife. Spain\\}
\affiliation{Departamento de Astrof\'{i}sica. Universidad de La Laguna. Tenerife. Spain\\}

\author{Beno\^{i}t Epinat}
\affiliation{Canada-France-Hawaii Telescope, 65-1238 Mamalahoa Highway, Kamuela, HI 96743, USA}
\affiliation{Aix Marseille Univ, CNRS, CNES, LAM, Marseille, France}

\author{Clare L. Dobbs}
\affiliation{School of Physics. University of Exeter, Stocker Road, Exeter EX4 4QL}

\author{Miguel Querejeta}
\affiliation{Observatorio Astron\'{o}nomico Nacional (IGN). Alfonso XII 3, Madrid. E-28014, Spain.}



\begin{abstract}

The Whirlpool Galaxy is a well studied grand design galaxy with two major spiral arms, and a large satellite NGC 5195. The arms both show long uniform sections with perturbations (“kinks” or sharp turns) in specific regions. Comparing the two arms shows a small radial offset between the main kinked regions. We analysed the morphology and also the velocity field in the disk of M51 using kinematic maps based on H$\alpha$ and CO line emission. These sample complementary radial ranges, with the CO map covering the central zone and the H$\alpha$ map extending to cover the outer zone. We looked for indicators of density wave resonance, zones where radial flows of gas in the disk plane reverse their sign. These were present in both velocity maps; their two-dimensional localization placed them along or closely parallel to the spiral arms, at a set of well defined galactocentric radii, and notably more concentrated along the southern, stronger arm. The results can be well interpreted quantitatively, using a numerical model of the interaction of M51 and NGC5195 in which the satellite has made two relatively recent passes through the disk plane of M51. During the first pass the pair of dominant spiral arms was stimulated, and during the second pass the strong kinks in both arms were formed at about the same time. The second interaction is particularly well characterised, because the timescale corresponding to the production of the kinks and the recovery of the original pitch angle is identical for the two arms.

\end{abstract}

\keywords{Galaxy spectroscopy(2171) --- Galaxy kinematics(602) --- Galaxy dynamics(591) --- Spiral arms(1559) --- Spiral pitch angle(1561) --- Galaxy interactions(600)}


\section{Introduction} \label{sec:intro}
Evidence for the presence of density waves in spiral galaxy disks has been presented in a numerous and wide variety of articles in the literature. In this article we will not attempt to summarize these publications, but will concentrate on a subset of those previous studies most relevant to the spiral structure in the disk of M51. There are half a dozen published methods for measuring pattern speeds and corotation radii of density wave patterns, each with their own limitations. These can be classified as direct methods and combined methods; the former are based entirely on observational data, such as the potential density phase-shift method \citep{buta_pattern_2009}, the Canzian method \citep{canzian_new_1993}, or the gravitational torque method \citep{querejeta_gravitational_2016}, while the later methods compare observational data with predictions of models or simulations \citep[see a detailed description in][]{font_ratio_2014}. Among all these methods, two have been used to produce the great majority of results. The classical, TW, method due to Tremaine and Weinberg \citep{tremaine_kinematic_1984} yields the pattern speed of a barred galaxy without requiring a dynamical model, measuring the radial velocity along a strip parallel to the line of nodes of the projected disk and the surface brightness of the component emitting the spectrum from which this velocity is derived. The method assumes that this component obeys a continuity equation, which means that strictly speaking it should be employed with the stellar population of the disk, and should be applied mainly to a stellar disk having a little internal absorption. The result of an integration using these measured parameters is a single value for the pattern speed averaged between the radial limits of the disk zone measured. For an assumed single disk pattern speed this radial extent should be as large as possible. After obtaining the pattern speed, the corotation radius can then be derived immediately from the observed rotation curve. The method was introduced for barred galaxies, where the dynamical effect of the bar is assumed to dominate, producing the rotational stationary wave in the disk beyond the bar. It has been generalized for more complex systems, in which different annular zones of the rotating disk may have different dominating pattern speeds \citep{meidt_radial_2008}, and various authors have used the TW technique with emission from interstellar gas using approximate arguments to justify the absence of continuity in the gas phase. A clear example, using H$\alpha$ emission from M100 was given in \cite{hernandez_bar_2004}.

An alternative method, to which we will refer as FB, was proposed by Font and Beckman \citep{font_resonant_2011}. It is based on the general prediction of the change in symmetry of orbits as we pass through corotation in a density wave \citep{kalnajs_confrontation_1978}. This change becomes evident in the kinematics of the streaming motions detectable in the gas of the spiral arms. At corotation the radial component of the streaming motion changes sign from inflow to outflow, or vice versa. Radioastronomers were aware of this prediction, and made use of it via kinematic maps of neutral atomic hydrogen at 21 cm to try to define a corotation radius in a limited number of nearby galaxies. \cite{vogel_h-alpha_1993} used a Fabry-Perot map of H$\alpha$ emission kinematics in M51, to detect and analyse the streaming motions along the spiral arms and showed that the radial components of these motions changed sign at well determined radii in the disk. Their work can well be considered a precursor to that of \cite{font_resonant_2011}, but they did not consider the possibility that a single density wave defined by a single corotation and its associated Lindblad resonances might not be a satisfactory basis for the full spiral arm structure. The FB technique is a generalization of the method used by \cite{vogel_h-alpha_1993}. The complete 2D Fabry-Perot kinematic map of a galaxy from its H$\alpha$ emission is analyzed as follows. Firstly a mean rotation curve is derived using all the data, taking due account of rotational geometry, and performing a harmonic fit. This curve is then projected azimuthally to give a 2D smooth mean velocity map of the galaxy, which is subsequently subtracted from the original map, to give a map of the (residual) non-circular velocities. Pixels are then identified where the radial velocity is zero within uncertainty. Each pixel is tested to find the change in value of the radial velocity as we cross the pixel in a radial direction in the disk. Only those where this change is more than twice the measurement uncertainty are maintained in the map. The final step is to divide the disk into annuli, and plot a 1d radial histogram of the numbers of these “zero” pixels in each annulus. The typical result, found in a couple of hundred galaxies, \citep[see e.g.][]{font_interlocking_2014}, is a histogram with a series of peaks. The galactocentric radius of each peak should correspond to a corotation, according to the predictions. In \cite{font_interlocking_2014} and a series of following papers \citep{font_ratio_2014,font_kinematic_2017, font_spiral_2019} the validity of this conclusion was tested. In this context special note should be taken of the work in \cite{beckman_precision_2018} on NGC 3433, a galaxy where it was practical to use both TW and FB, with the relevant kinematic and photometric data from stars and also from interstellar H$\alpha$. Both methods yielded a dominant pattern speed and corotation, with the same values, well within the error limits of each, derived separately.  Four detailed data sets could be used with the TW and FB methods using both stellar and interstellar data. However FB yielded three further corotations, one corresponding to an interior radius, and two placed beyond the dominant corotation, out in the spiral arms. Of these, only the outermost was also obtainable using TW, with kinematic data from the gas. The overall conclusion is that FB, where the galaxy has sufficient star formation in the disk to produce disseminated H$\alpha$ emission, yields a more complete analysis of the resonant structure of a disk, where it can be compared directly with TW, gives the same result.

In previous studies using FB the one-dimensional radial histograms have given the information needed to establish the galactocentric radii of the corotations. But the more detailed relations of the siting of the zero radial velocity pixels in relation to morphological features were not studied. In the present article we take advantage of the high spatial and spectral resolution in a Fabry-Perot map of M51 to examine these relations with respect to the spiral arm morphology.

\section{The observations of M51} \label{sec:data}

The Whirlpool galaxy, M51, is a nearby grand design galaxy with quite low inclination angle, which makes this galaxy a preferred target of numerous observations and numerical models, which have led to hundreds of published studies. Table \ref{tab:parameters} shows the main parameters of M51, as well as the sources from where the values are obtained. This galaxy has been observed in all possible wavelengths of the electromagnetic spectrum, from radio waves up to the X-ray domain, including $\gamma$-rays, making the Whirlpool galaxy a real multimessenger source. In the present study we combine precise observations of the ionised gas, mapping the H$\alpha$ emission line all over the whole galaxy, with precise observations of the molecular gas, tracing the CO(1-0) emission line in the central region of the galaxy. The kinematic position of the center of M51 is found to be the same for the two sets of data, and so is the inclination angle; however, there are two position angles of the major axis (which do not coincide) for the H$\alpha$ data and the CO data, as indicated in Table \ref{tab:parameters}. This difference is quite small, and can be attributed to the azimuthal offset of the spiral arms in CO with respect to those in H$\alpha$, as discussed in the following section. The values calculated here are in agreement with those determined in previous studies \citep[see table 1 in][]{colombo_pdbi_2014}. 

\begin{table}[h!]
    \centering
        \caption{Parameters of M51}
            \label{tab:parameters}
    \begin{tabular}{c c c}
    \hline \hline
    Parameter & Value & Source\\
     \hline
    Morphological type & S\underline{A}B(rs,rs)bc &  \cite{buta_classical_2015}\\ 
    kinematic center RA (J2000)  &  13h:29m:52.9s   &  This study\\
    kinematic center DEC (J2000) &  +47d:11m:43.2s & This study \\
    Systemic velocity & 470.5$\pm$1.6 km~s$^{-1}$  &  This study\\
    Inclination Angle & $22^{\circ}$ & This study \\
    Position Angle & $169^{\circ} (H\alpha), 175^{\circ} (CO)$ & This study \\
    Distance & 8.34 Mpc  &  NED databse \\
\hline
    \end{tabular}
\end{table}

\subsection{The ionised gas} \label{subsec:Halpha}
Fabry-Perot data of M51 (NGC5194) in the wavelength range around the H$\alpha$ line were observed in May 2003 at Observatoire du Mont Mégantic with the FaNTOmM instrument as part of the SINGS survey \citep{daigle_hkinematics_2006}. These data were acquired with a photon-counting camera with $512\times 512$ 1.61\arcsec\ pixels covering an unvignetted field-of-view of $13.74\times 13.74$ square arcminutes, which corresponds to $33.3\times33.3$ kpc of M51, and contains 48 spectral channels each of width  7~km~s$^{-1}$ obtained with a scanning Fabry-Perot interferometer having a spectral resolution $R=18609$ at the observed wavelength of the H$\alpha$ line, for a total exposure time of 4.1 hours (5.1 minutes per channel).

During the data reduction process performed by \cite{daigle_hkinematics_2006} on M51, the datacube was first wavelength calibrated, then sky-subtracted, and finally either smoothed or binned adaptively. To gain more control over the data analysis, we use the sky subtracted datacube before smoothing or binning. We perform our own 3 pixels FWHM Gaussian smoothing on each slice of the cube to increase slightly the signal-to-noise ratio (SNR) while hardly affecting the spatial resolution which after processing is $\sim 5.8$ \arcsec\ FWHM.

Since scanning Fabry-Perot data is periodic in the spectral dimension, we use the same dedicated algorithm as that used in the original data reduction \citep{daigle_hkinematics_2006} to compute line moments from barycenter measurements for each pixel and derive maps in the continuum, dispersion around continuum, H$\alpha$ line flux, radial velocity and velocity dispersion maps. Finally, we create masks in order to identify spurious pixels with low SNR that have velocity values incoherent with the large scale pattern. We use a new automatic procedure for this task that segments the velocity field based on velocity field continuity, and sorts the segments based on their SNR.

\begin{figure}[ht]
    \centering
    \plotone{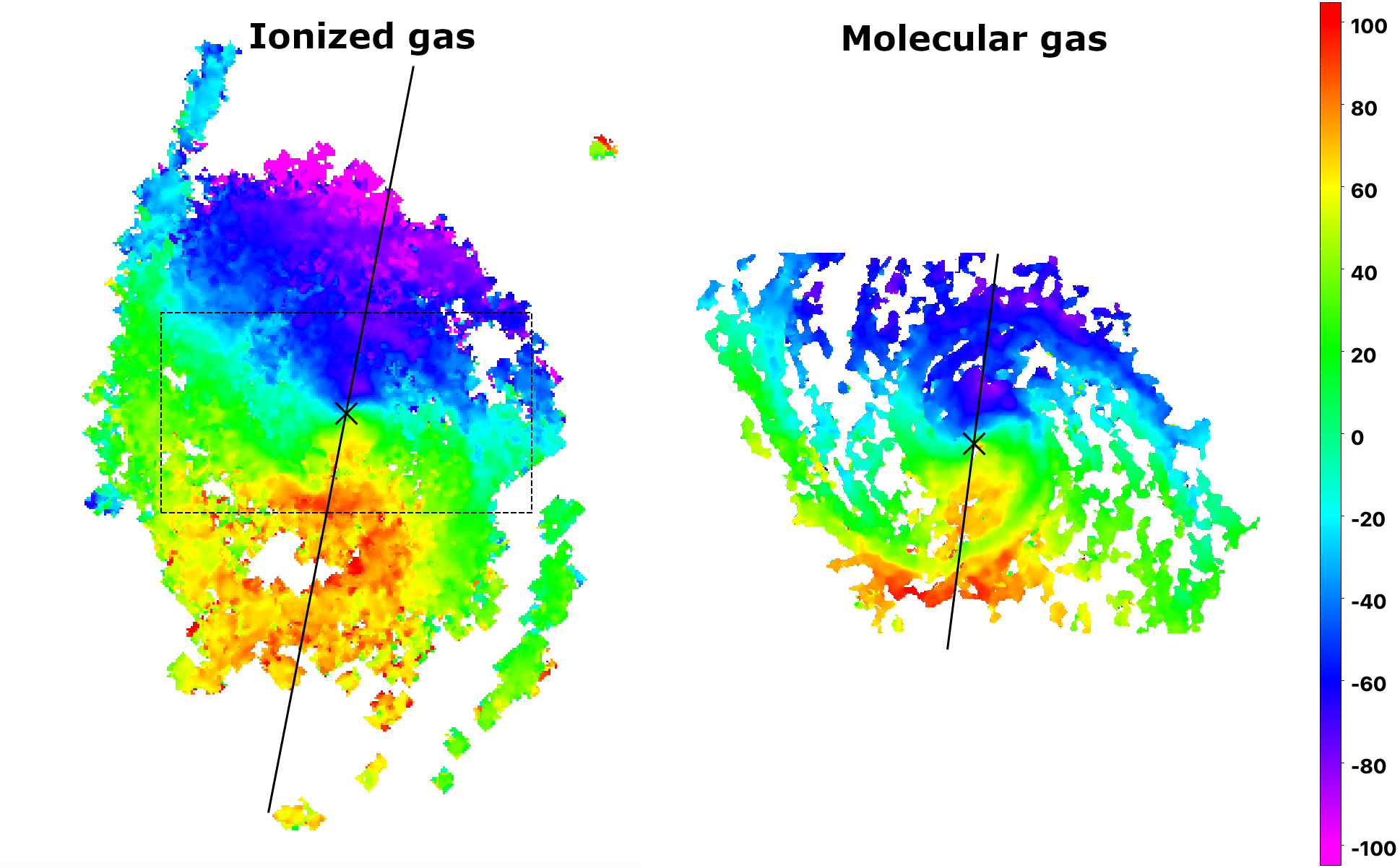}
    \caption{Velocity map derived from the $H\alpha$ data-cube in the left panel, and from the CO(1-0) data-cube in the right panel. The color scale to the right side indicates the velocity in ~km~s$^{-1}$. The kinematic center is marked with an $\times$ and the kinematic Position Angle is shown with a solid line in black. The dashed box plotted over the ionized gas velocity map shows the field covered with the molecular gas data. North is up and East is left in both maps.}
    \label{fig:velmap}
\end{figure}

The left panel of Figure \ref{fig:velmap} shows the velocity map extracted from the H$\alpha$ observations, and in the right panel, the equivalent map obtained from the CO observations, as detailed in section \ref{subsec:CO} below.

\subsection{The molecular gas} \label{subsec:CO}

We also work with an alternative tracer, that of the molecular ISM. We employ the CO(1-0) intensity maps from the PdBI Arcsecond Whirlpool Survey (PAWS; \cite{schinnerer_pdbi_2013}), which traces the bulk molecular gas in the central ${\sim}11 \times 6$\,kpc of M51. The observations were taken with the Plateau de Bure interferometer (PdBI) in all configurations between August 2009 and March 2010, and were corrected for missing short spacings with single-dish observations from the IRAM 30\,m telescope (obtained in May 2010). The data reduction is described in detail in \cite{pety_plateau_2013}. The resulting spatial resolution is ${\sim}1''$, but here we use a version tapered to $3''$, which produces a higher SNR. Here we consider the moment-0 intensity map, produced using a combination of masks \citep[dilated mask and HI prior; see Appendix B in][]{pety_plateau_2013}. We transformed the measured CO(1-0) intensities into molecular hydrogen gas surface densities (H$_2$) using a constant Galactic conversion factor, $\alpha_{\rm CO} = 4.35$\,M$_\odot$\,pc$^{-2}$\,(K\,km\,s$^{-1}$)$^{-1}$ \citep{bolatto_co--h2_2013}. Any radial variations in $\alpha_{\rm CO}$ should have no significant consequences for our method, as we consider azimuthal profiles.
The cubes have channels of 5~km~s$^{-1}$, and here we consider a velocity map obtained as a 1st-order moment map. In order to produce the moment map, emission in the CO cube is first masked according to a combination of a dilated mask (based on adjacent high S/N regions) and an HI velocity prior \citep[see Appendix B in ][]{pety_plateau_2013}.

\section{Morphology of M51} \label{sec:pitch}

\begin{figure}
    \centering
    \plotone{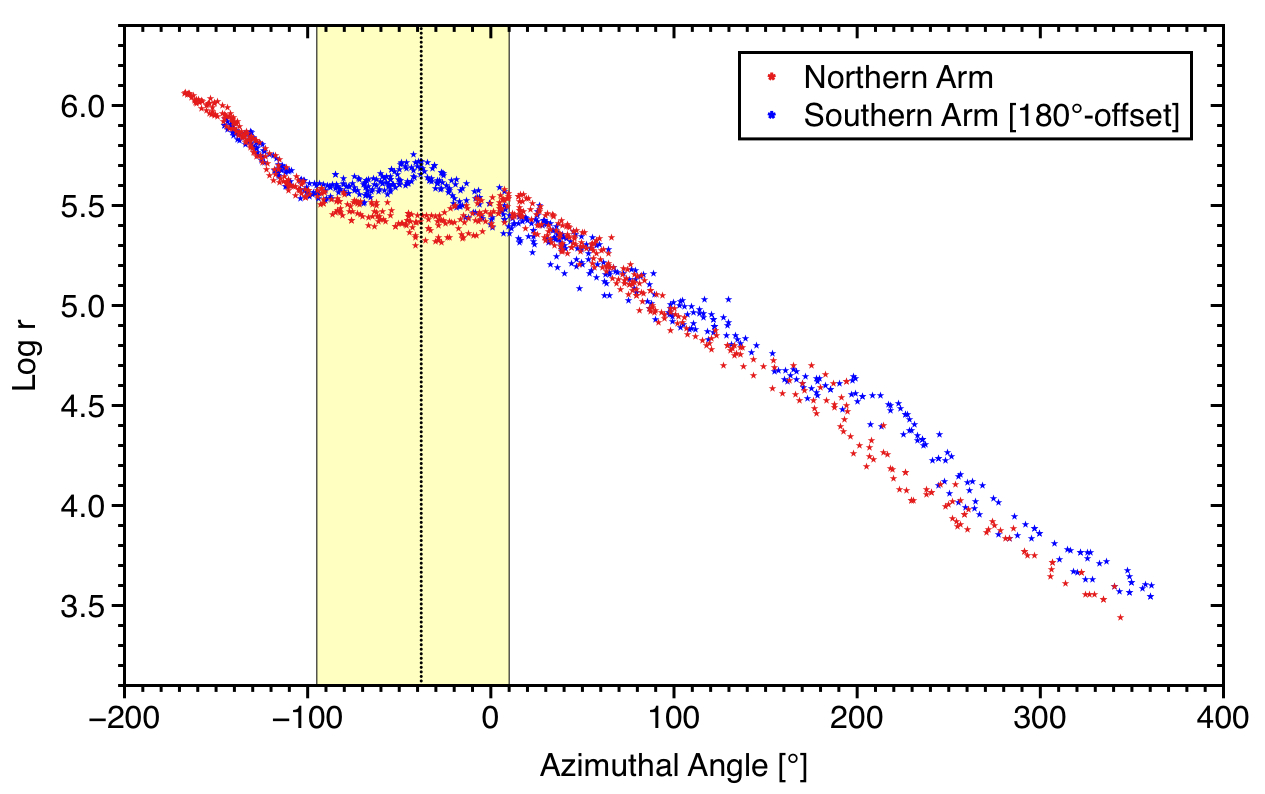}
    \caption{Plot of the two spiral arms of M51 in the ($\log{(r)},\theta$) plane (with \textit{r} in pixel units), the Northern arm in red and the Southern arm in blue. An offset of $180^\circ$ is applied to the Southern arm. The vertical shaded region in yellow indicates the perturbed region, (the 'kink' region) and the vertical line marks the position of the kink of the Southern arm.}
    \label{fig:N&S}
\end{figure}

\begin{figure}
    \centering
    \plotone{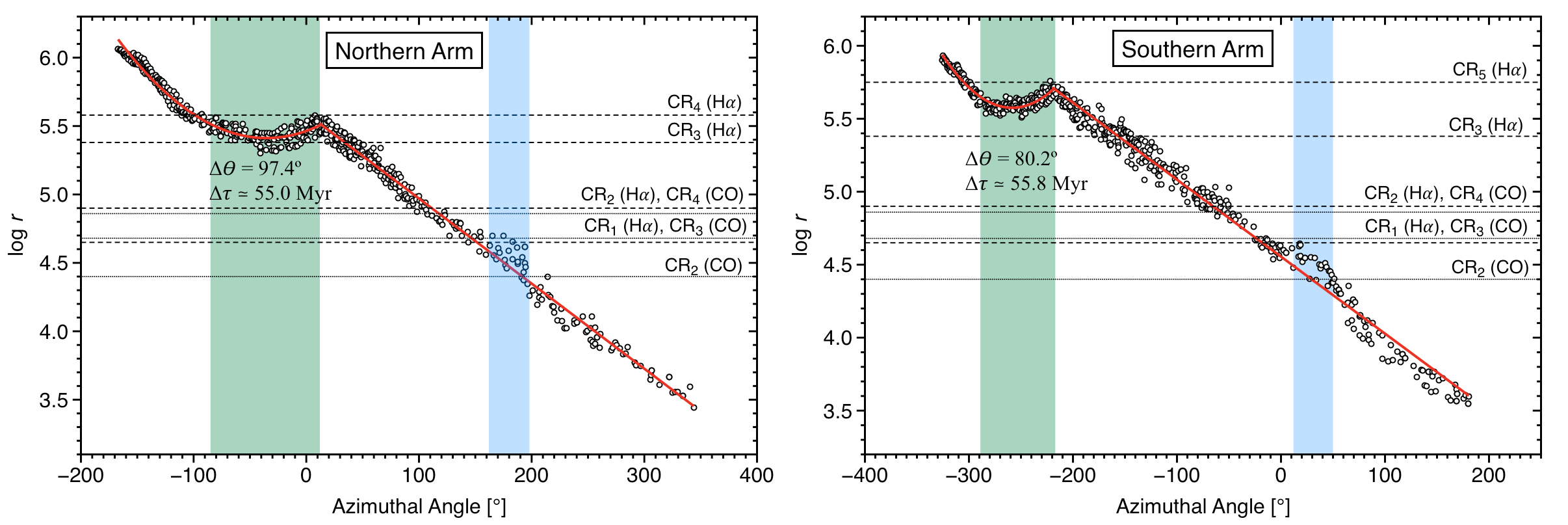}
    \caption{\textit{Left}. Plot of the Northern arm of M51 in the ($\log{(r)},\theta$) plane (with \textit{r} in pixel units). The shaded region in green shows the perturbation region, which covers an azimuthal range $\Delta\theta$, is characterised with $\Delta\tau$ (see section \ref{sec:resonances} for a detailed description). The different fits are also plotted in red. Additionally, the H$\alpha$ and the CO corotation radii associated with this arm are also indicated as horizontal dashed and dotted lines, respectively, and are labelled accordingly \textit{Right}. The same as left panel but for the Southern arm.}
    \label{fig:armfit}
\end{figure}

The ideal coordinate system for our analysis is $(\log{(r)},\theta)$, where $r$ is the galactocentric radius and $\theta$ is the azimuthal angle. In this system, logarithmic spirals are linear, and the slope of the linear fit gives the pitch angle, $\varphi$. The 3.6 $\mu$m Spitzer image was used for this morphological work, in order to minimize any geometrically distorting effects of dust, and the spiral arms were delineated by distributing pixel points on the arms wherever the surface brightness exceeded a predetermined lower limit. The results of these plots are shown in Fig. \ref{fig:N&S} in which the northern and southern arms are plotted together. Here, the northern arm is the one pointing to the companion galaxy NGC 5195, which is in the North of the image (see the Spitzer image in Fig. \ref{fig:PR2DSpitzer}). These graphs have the virtue of showing clearly the range of a spiral arm along which the pitch angle is virtually constant, and identifying very clearly the points where the pitch angle changes. There is a striking quantitative morphological similarity between the two arms: both show perturbed regions where the arms are “kinked” back inwards towards the centre of the galaxy over well defined ranges, and then resume their outward spirals with pitch angles very close to the original values interior to the kinks. The graphs show that there is a separation in angle of $\pi$ between the two arms, thus even with the kinks, M51 is a good example of a grand design spiral galaxy. In Fig. \ref{fig:N&S} the yellow shaded area marks the range of angles between which the kinked sections of the arms are found, which we term the “kink region”. Although the overall offset between the two arms is just $\pi$ there is a subsidiary offset for the kinks: the southern arm reaches a local maximum radial distance from the centre at an angular distance of $\pi$ from where the northern arm reaches a local minimum. This is marked with a vertical solid line in Fig. \ref{fig:N&S}, and there is a $\sim50^\circ$ incremental offset between the local maxima and also between the local minima for the two arms. As for the radial positions of the kinks, the northern arm kink is found at a radial distance of 5.9 kpc, while the kink in the southern arm is located at 7.2 kpc. A closer look of Fig. \ref{fig:N&S} reveals that there is a weak kink feature in the inner region at a distance of 2.6 kpc from the center of the galaxy in both spiral arms. This small perturbation is clear in the southern arm, but also noticeable in the northern arm (see left and right panels of Fig. \ref{fig:armfit}) where this feature is indicated with a blue shaded vertical region. A careful inspection shows a further small difference between the arm morphologies: the southern arm has a slightly smaller pitch angle than the northern arm. In the $\pi$-shifted plot it trails the northern arm in the inner section of the disc, then crosses the northern arm to lead it at increasing radius until the kink region.

For a quantitative treatment of the arms we have performed functional fits to their defined portions. These fits are shown in Fig. \ref{fig:armfit} for the two arms. In both arms a single linear fit gives a good global description of the inner sections of the arm out as far as the kink, allowing a single pitch angle to be derived for each arm in those sections. The values for these pitch angles are given in Table~\ref{tab:pitchangleM51}. Beyond the kink regions, it is clearly preferable to make a second order polynomial fit, and these fits are shown in Fig.\ref{fig:armfit}. The kink regions are shaded in green for the two arms. The corotation radii associated with the arm are marked in each plot with an horizontal dashed line for the H$\alpha$ resonances and with horizontal dotted line for the CO resonances (see section \ref{sec:resonances} for a detailed description of the resonances); we label the corotations $CR_1$(H$\alpha$/CO), $CR_2$(H$\alpha$/CO), $\ldots$, in agreement with the resonances found in H$\alpha$ or CO, which are shown in Fig. \ref{fig:histograms}. In order to examine the behavior of the pitch angles, however, we treated the kink regions in an alternative way, making asymptotic linear fits to the zones right after the kinks, as far as the pitch angle minimum $\varphi_{rev}$ and measuring the deviation, $\Delta\varphi_{kink}$, for each arm, of the pitch angle between this section and the inner section . The pitch angles of the outer sections of the arms can be measured using single linear fits in each arm, and the difference between the pitch angle of the inner and outer linear sections of the arms, $\Delta\varphi_{out}$ can be measured for each arm. It is interesting to note that the values of $\Delta\varphi_{kink}$ and $\Delta\varphi_{out}$ respectively are closely similar in both arms. 

The pitch angle of the spirals of M51 has been estimated in a variety of studies using different types of observations. Table \ref{tab:pitchangle} summarizes the results found in previous studies, along with the type of observations used to derive the pitch angle. In most of these studies, a single value of the pitch angle is provided, which is consistent with the pitch angle we determine for the inner region,  $\varphi_{main}$  in Table~\ref{tab:pitchangleM51}. We find particularly interesting the analysis of \cite{egusa_gas_2017}, where gas and stellar data are used to derive the pitch angle of the inner regions for each arm separately of the main regions, included in Table~\ref{tab:pitchangle}, but the authors also determine the pitch angle in the outer region for each arm, obtaining $27.8^{\circ}\pm 1.3^{\circ}$ for the gas component, $37.4^{\circ}\pm 3.5^{\circ}$ for the stellar component of the southern arm, and similarly, $27.6^{\circ}\pm 1.5^{\circ}$, $32.4^{\circ}\pm 2.8^{\circ}$ for the northern arm; these values are clearly larger, in particular for the stellar component, than the values of $\varphi_{out}$ given in Table~\ref{tab:pitchangleM51}. The most detailed analysis of the spiral arms in terms of pitch angle was performed by \cite{honig_characteristics_2015}. In that study, the authors fragment the arms into several azimuth ranges and determine the pitch angle in each segment \cite[see table 4 in][]{honig_characteristics_2015}. Following this method, the authors find a segment in each arm in which the pitch angles reverse their signs, and thus they identify the kink in each arm with a radial position compatible with the positions we determine in the present study.

Another morphological feature that is considered here is the nuclear bar of M51. \cite{buta_classical_2015} (see Table~\ref{tab:parameters}) have classified this galaxy as barred, and \cite{herrera-endoqui_catalogue_2015} using \textit{Spitzer} images have estimated a deprojected radius of $r_{bar}=22.1$\arcsec\ , which is consistent with the bar length of $20.1$\arcsec\ determined in \cite{comeron_ainur:_2010}, who used archival data from the \textit{Hubble} Space Telescope.

\begin{table}[ht]
    \centering
       \caption{Variation of the Pitch Angle in M51}
    \label{tab:pitchangleM51}
    \begin{tabular}{c c c c c c}
    \hline \hline
      {    } & Main region   &  reverse region  & Change at kink & Outermost region & Change after kink  \\
      {}  &  $\varphi_{main}$ & $\varphi_{rev}$ & $\Delta\varphi_{kink}$ & $\varphi_{out}$ & $\Delta\varphi_{out}$\\
      \hline
      N Arm  & $19.6^{\circ} \pm 0.4^{\circ}$ & $-7.4^{\circ} \pm 0.9^{\circ}$ & $27.0^{\circ} \pm 1.0 ^{\circ}$ & $25.4^{\circ} \pm 0.6^{\circ}$ &  $32.8^{\circ} \pm 1.2 ^{\circ}$ \\
      S Arm  & $16.8^{\circ} \pm 0.4^{\circ}$ & $-12.0^{\circ} \pm 1.4^{\circ}$ & $28.8^{\circ} \pm 1.5 ^{\circ}$ & $22.0^{\circ} \pm 0.6^{\circ}$ &  $34.0^{\circ} \pm 1.5 ^{\circ}$ \\
      \hline
    \end{tabular}
\end{table}

\begin{table}[ht]
    \centering
        \caption{Pitch Angle of M51 Spiral Arms}
    \label{tab:pitchangle}
    \begin{tabular}{c c c}
        \hline \hline
     Reference    & Values  &  Data \\
     \hline
         \cite{elmegreen_spiral_1989} & $15^{\circ} \pm 4^{\circ}$ & B-band\\
         \cite{nakai_distribution_1994} & $21^{\circ} \pm 5$ & CO(J=1-0)\\
         \cite{shetty_kinematics_2007} & $21.1^{\circ}$ & CO(J=1-0)\\
         \cite{fletcher_magnetic_2011} & $20^{\circ}$ & $\lambda20cm$\\
         \cite{puerari_new_2014} & $19^{\circ}$ & 8 $\mu$m\\
         \cite{miyamoto_influence_2014} & $19^{\circ} \pm 1^{\circ}$ & CO(J=1-0)\\
         \cite{egusa_gas_2017} & $18.8^{\circ}\pm 0.6^{\circ}$, $23.4^{\circ}\pm 0.9^{\circ}$ (N Arm) & CO(J=1-0), Stellar\\
         {}  & $19.9^{\circ}\pm 0.3^{\circ}$, $19.3^{\circ}\pm 0.5^{\circ}$ (S Arm) & \\
         \cite{brok_co_2022} & $20^{\circ}$ & several lines \\
         This study &  $19.6^{\circ} \pm 0.4^{\circ}$ (N Arm) & 3.6 $\mu$m \\
         {} & $16.8^{\circ} \pm 0.4^{\circ}$ (S Arm) & \\
    \hline
    \end{tabular}

\end{table}

\section{Resonances of M51} \label{sec:resonances}

\begin{figure}
    \centering
    \plotone{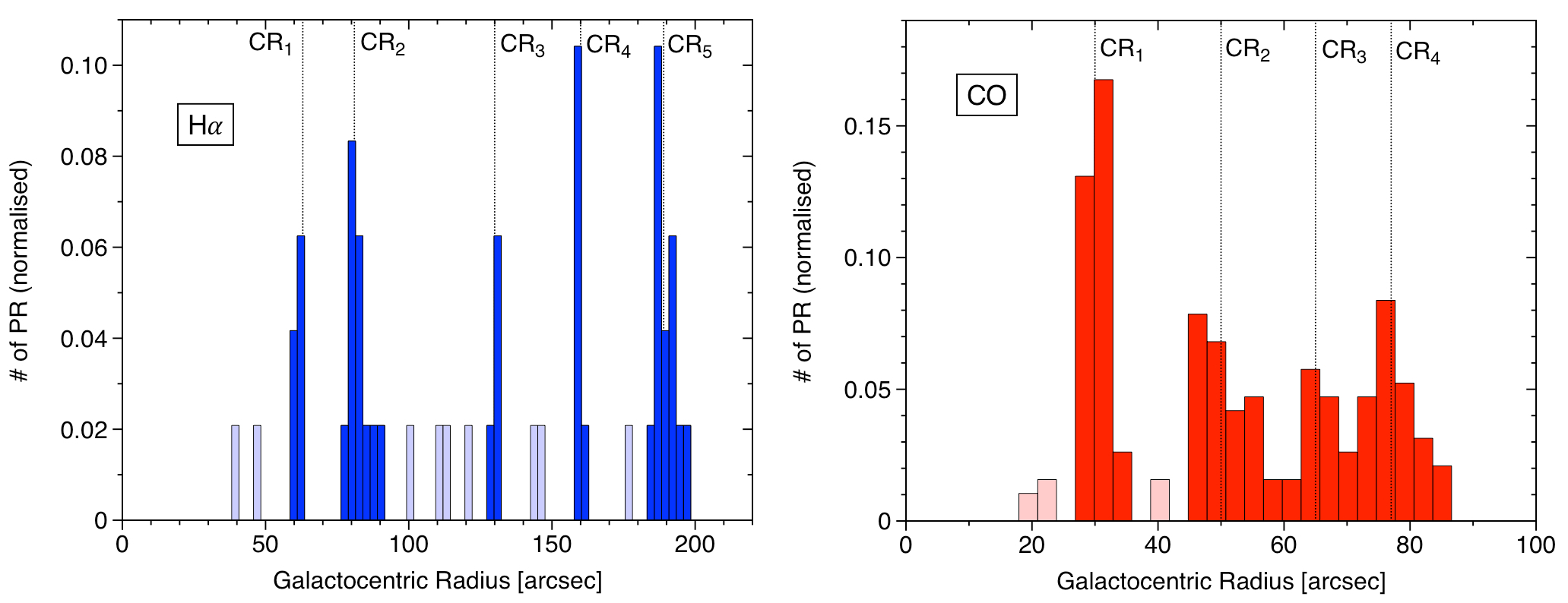}
    \caption{\textit{Left panel}. Radial distribution of the Phase Reversals obtained with the H$\alpha$ velocity map. The peaks in solid color indicate the corotation radius of the resonances found, which are marked as vertical dotted lines and are labelled as $CR_1$, $CR_2$, $\ldots$ The shaded weak peaks are not relevant and do not contribute to any resonance. \textit{Right panel}. The same as left panel, but for the molecular gas.}
    \label{fig:histograms}
\end{figure}

The FB method used to find the resonance radii and pattern speeds in M51  has been described previously in detail \citep{font_resonant_2011,font_interlocking_2014} so here we just describe it quite briefly. We first note that in \cite{beckman_precision_2018} we showed, with a meticulous treatment of the internal dynamics of NGC 3433, that FB produced identical values for the main bar pattern speed and corotation radius to those obtained using the “classical” TW method, and also identified additional corotations. The technique uses a 2D mapping of the internal kinematics of a galaxy disc to find the zones with a concentration of pixels where the radial velocity in the disc plane changes sign (named phase reversals), from inflow to outflow or vice versa. The final product of the method is the radial distribution of the phase reversals, which can be plotted as a histogram, resulting in a series of peaks each of which defines a corotation radius. The FB method was performed in the present study on the H$\alpha$ map and also on the CO map of M51. The normalised histograms derived from both maps are shown in Fig. \ref{fig:histograms}, where we have marked the position of the corotation radii as vertical dotted lines. We can see in Fig. \ref{fig:histograms} that the CO map extends out to radius of $\sim$ 120\arcsec\, but has full azimuthal coverage only out to $\sim$ 80\arcsec\ from the center, so we cannot use it to derive corotations or pattern speeds beyond that radius. However the excellent angular resolution provided by PdBI allows us to identify four corotations within this range. The complementarity between the two data sets is shown by the fact that although the H$\alpha$ velocity map does not show resonances for galactocentric radii smaller than $\sim$ 50\arcsec\ the map extends out to $\sim$ 200 \arcsec\ from the center. We find four corotations using the CO map, and five resonance radii using the H$\alpha$ map, of which the two innermost are consistent, within the uncertainties, with the two outer CO resonances (see Table \ref{tab:resonances} for a quantitative comparison).

\begin{figure}
    \centering
    \plotone{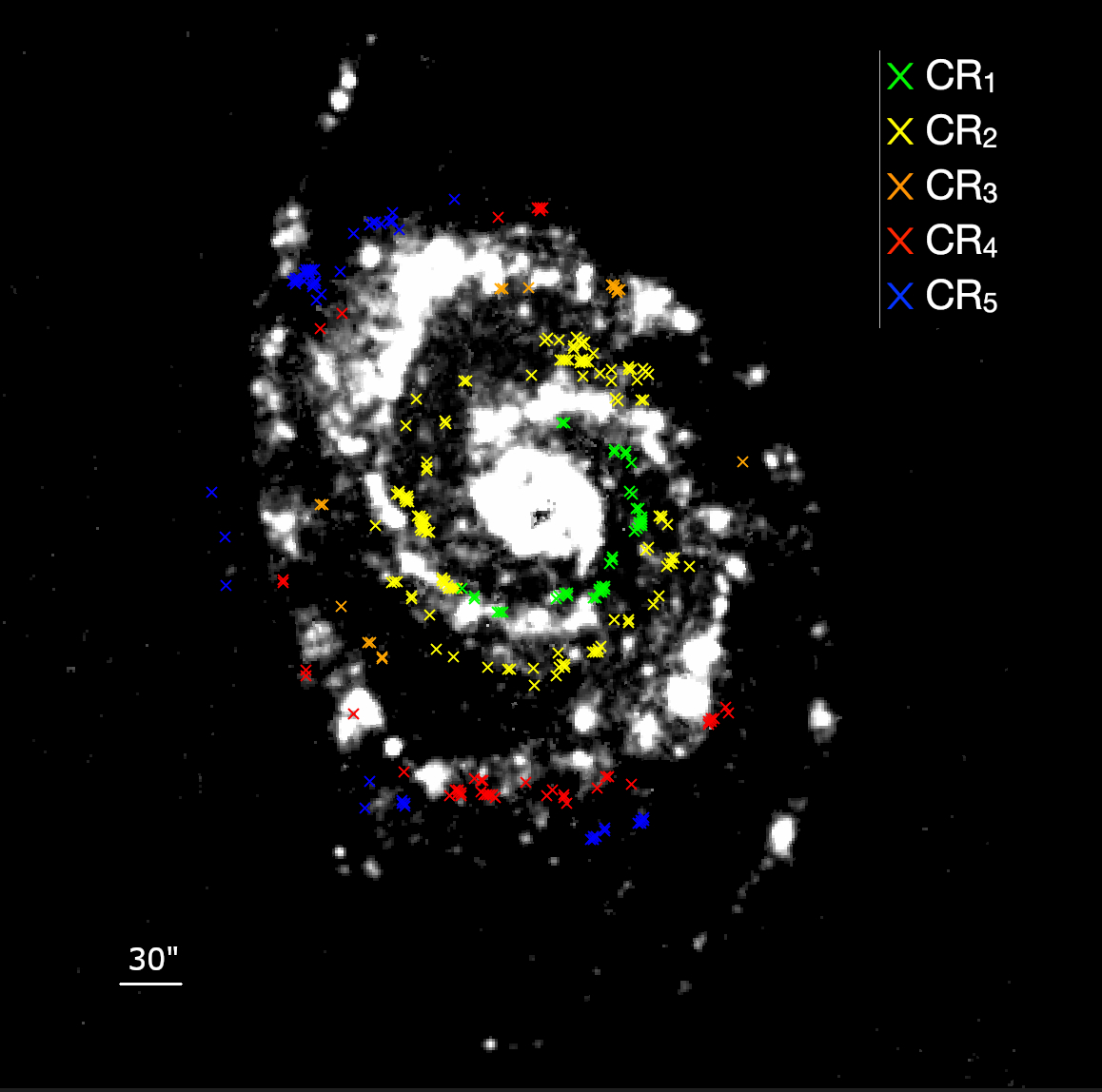}
    \caption{Phase reversals identified from the Fabry-Perot data are overplotted on the $H\alpha$ intensity map. Each color indicates a different peak of the histogram shown in Fig.\ref{fig:histograms}. North is up and East is left.}
    \label{fig:PR2DHa}
\end{figure}

\begin{figure}
    \centering
    \plotone{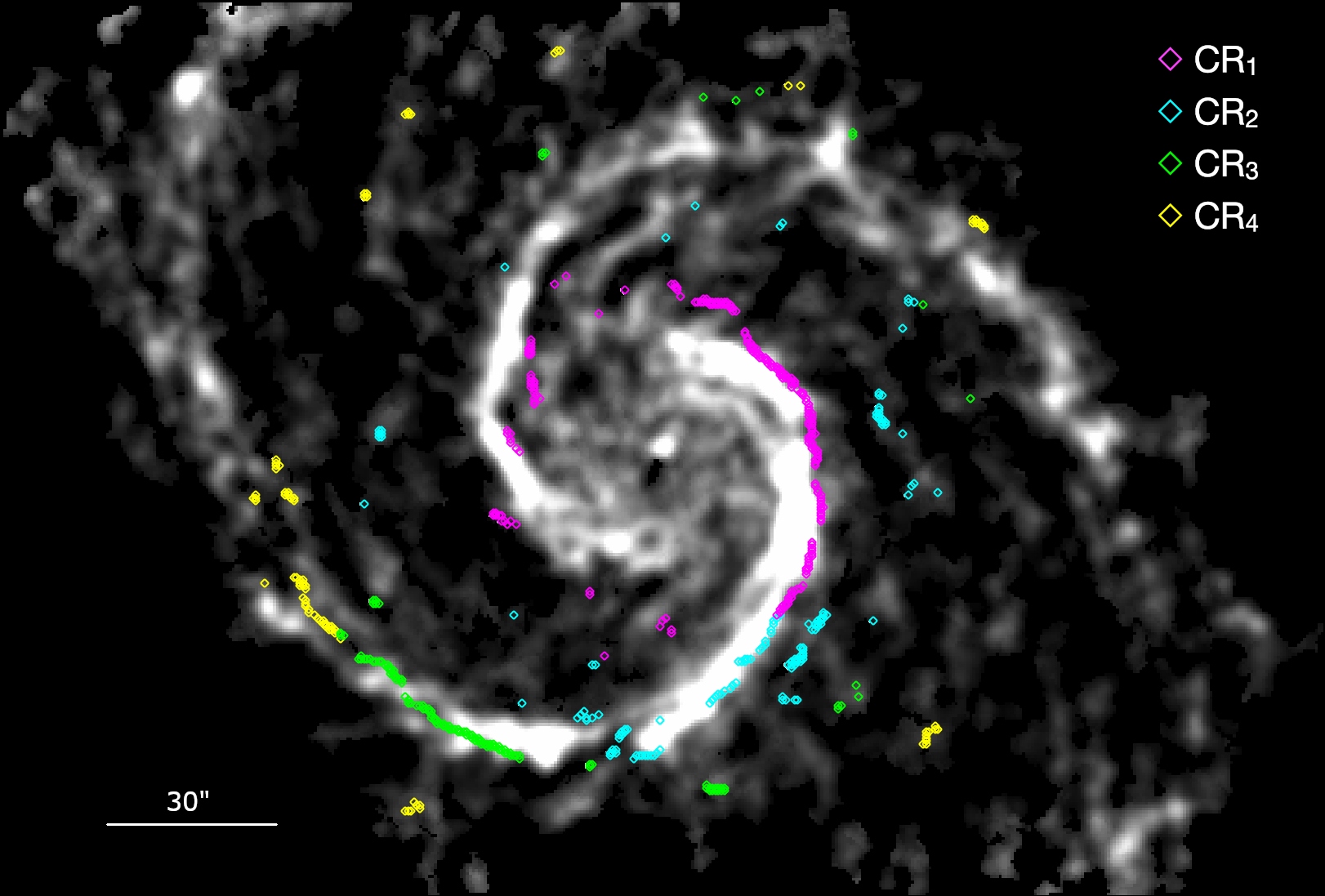}
    \caption{Phase reversals found from the CO data overplotted on its corresponding moment zero map of CO. Each color indicates a different peak of the histogram shown in Fig.\ref{fig:histograms}. North is up and East is left.}
    \label{fig:PR2DCO}
\end{figure}

\begin{figure}
    \centering
    \plotone{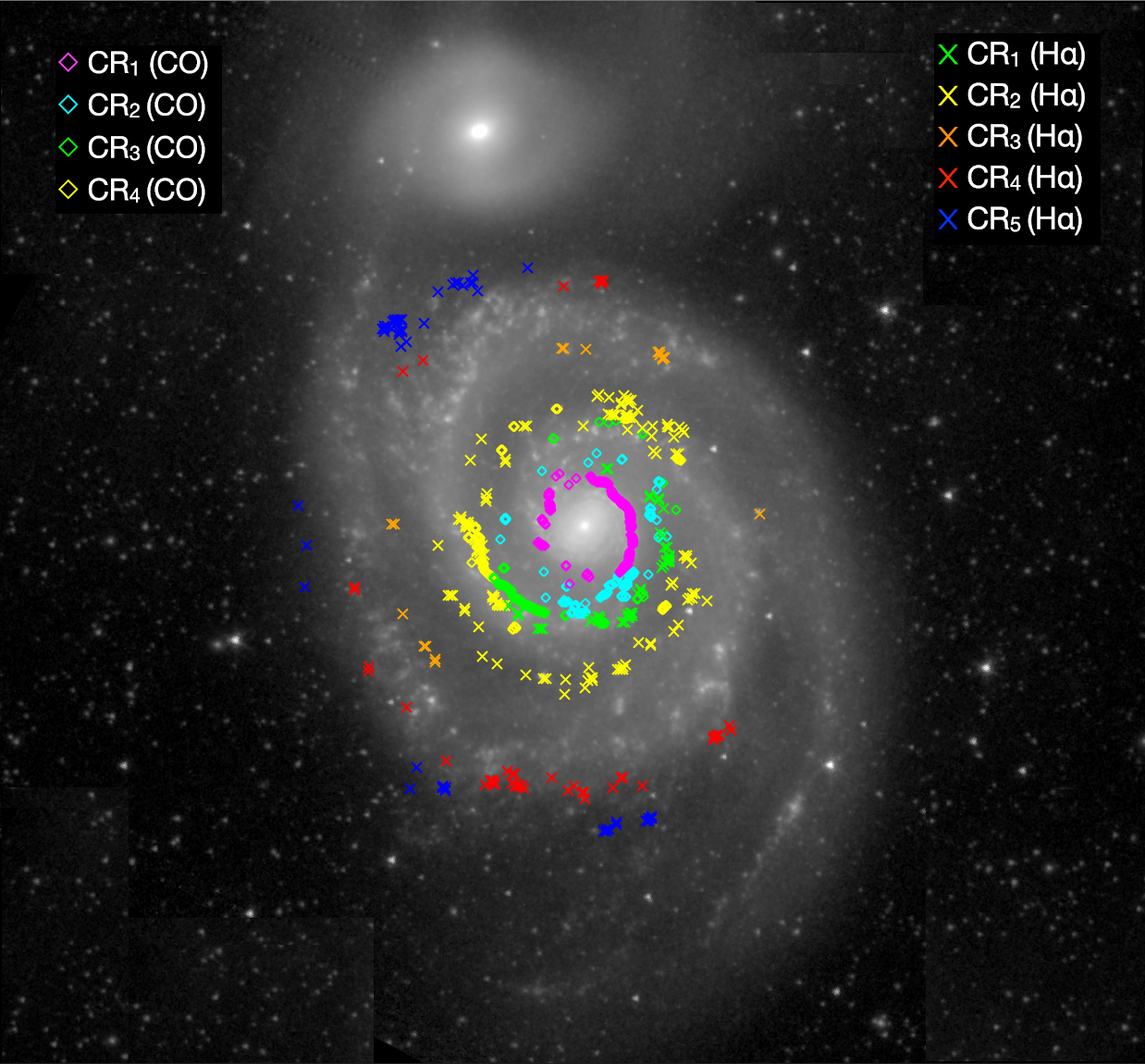}
    \caption{The $H\alpha$ and CO  phase reversals, marked with $\times$ symbol and diamond, respectively, displayed on the Spitzer image of M51. The resonances shown in Figure \ref{fig:histograms} are coded in different colors as indicated in the legends. North is up and East is left.}
    \label{fig:PR2DSpitzer}
\end{figure}

In order to obtain a better understanding of the observed kinematic distributions we have plotted the points where the radial velocity flow changes direction from inwards to outwards on a 2D image of M51. To do this we relaxed our previously established condition for the assignment of these phase reversal points. Instead of selecting  only those points with zero radial velocity where the change when going from larger to smaller radii is at least as big as the measured uncertainty in the mapping velocity, we have enlarged the number of points by including those where the change is greater than one half of the velocity uncertainty. The phase reversal points are overplotted on surface brightness maps. In Fig. \ref{fig:PR2DHa}, the H$\alpha$ phase reversals are shown on a map of H$\alpha$ surface brightness; each resonance identified in Fig. \ref{fig:histograms} is plotted with a different color. In Fig. \ref{fig:PR2DCO}, the CO phase reversals are shown on the zeroth moment map in CO, similarly, the four resonances found in CO, shown in the histogram of Fig.\ref{fig:histograms}, are displayed with different colors. Note that the angular scale is indicated in the two images with a bar of 30 \arcsec length. All the phase reversals, for the H$\alpha$ with $\times$ symbols and for the CO with diamond symbols, are plotted on a Spitzer 3.6 $\mu$m continuum image in Fig. \ref{fig:PR2DSpitzer}, which shows that the first corotation in H$\alpha$ overlaps with the third corotation found in CO (the two of them are plotted in green but with different symbols). The same situation happens with the second corotation in H$\alpha$ and the fourth resonance in CO (in yellow in Fig.\ref{fig:PR2DSpitzer}).

In a number of previous articles the existence of more than a single pattern speed with its associated corotation radius has been suggested and demonstrated for disk galaxies \citep{meidt_radial_2008, font_interlocking_2014, font_spiral_2019, beckman_precision_2018}. For barred galaxies the dominant density wave is that associated with the bar, but there is usually more than one associated with the spiral arms. In M51 \cite{elmegreen_spiral_1989} had already suggested that the morphology of the galaxy cannot be well explained with a single pattern speed, and \cite{dobbs_simulations_2010}, claimed that each arm should have its own radially varying pattern speed. \cite{meidt_radial_2008}  confirmed the prediction of multiplicity, finding three pattern speeds within the disk. Our results presented in Fig. \ref{fig:PR2DSpitzer} show that the phase reversals are strongly associated with the spiral arms, responding to different density waves for each arm. This result is in agreement with the hydrodynamical simulations by \cite{baba_gas_2016}, and supports the dynamical spiral theory, which describes the spiral arms as differentially rotating, transient but recurrent patterns \citep{dobbs_dawes_2014}. The quantities for the relevant parameters are collected in Table \ref{tab:resonances} where we present the corotation radii we derive using FB, and the corresponding pattern speeds calculated assuming the value for the distance to M51 used within the present study. We also offer in Table \ref{tab:patternspeed} the results of the pattern speed corrected for an assumed distance of 8.34 Mpc used in other studies in the literature, for comparison with the values found in the present study. In the last column of Table \ref{tab:resonances} the morphological feature associated with each corotation is defined. To make our results clear, here is the explanation for each of these associations.

$CR_4$ (H$\alpha$) and $CR_5$ (H$\alpha$) are very close to the kinks making a virtually tangential contact with the kinks in the two arms, as shown in Fig. \ref{fig:armfit}. $CR_5$(H$\alpha$) almost touches the kink in the southern arm, and $CR_4$ (H$\alpha$) is very close to the kink in the northern arm. We interpret this as giving us a criterion for associating each resonance with its corresponding arm. Using the azimuthal range of the region of the kink perturbations,  $\Delta\theta$ (the green shaded regions in Figure \ref{fig:armfit}) and the pattern speed associated with each arm in its respective azimuthal range, ($\Omega_4$ for the northern arm and $\Omega_5$ for the southern arm) we can estimate the timescales for the transmission of the perturbations. These are calculated to be $\Delta\tau_N$ = 55.8 ($\pm$2.1) Myr for the northern arm, and $\Delta\tau_S$ = 55.0 ($\pm$1.1) Myr for the southern arm. These timescales are the same, which strongly suggests that the source of the perturbation which caused the kink and is then propagated through the perturbation region was the same for both arms.

$CR_1$ (H$\alpha$) and $CR_3$ (CO) have a similar corotation radius (see table \ref{tab:resonances}), meaning that the same resonance is found for the ionised gas and the molecular gas. Fig. \ref{fig:armfit} shows that this resonance is very close to the weak kink, which is peaked at $r\sim 65$\arcsec, so following the previous criterion we can associate this resonance with the weak kink, which is present in both arms.

$CR_1$ (CO) is the strongest resonance feature, and is caused by the bar. The rotational parameter, $\Gamma$  which is defined as the ratio of the corotation radius to the bar length, $\Gamma$ is measured as 1.48. According to the widely used criterion of \cite{debattista_constraints_2000} this would make the bar a slow rotator, but its pattern speed, 135 $kms^{-1}kpc^{-1}$ is really fast. This result is compatible (within the error limits)  with that of \cite{meidt_radial_2008} who used a radial variant of the TW method, but significantly smaller than a value by \cite{querejeta_gravitational_2016} using a method of torques. We should point out that in a previous article \citep{font_spiral_2019} we explain in detail why the Debattista and Sellwood criterion should not be generally applied.

$CR_2$ (H$\alpha$) and $CR_4$ (CO) are located in the same radial position, taking into account the uncertainties, thus we can consider that we are dealing with a single dynamical resonance, which cannot be related to any feature of the spiral arms, such as kinks or bifurcations. Therefore, this resonance is associated with the two arms. The same situation happens with two other resonances: $CR_2$ (CO) and $CR_3$ (H$\alpha$); each of them has its own corotation radius and pattern speed, and we consider that they are also associated with the spiral structure of the galaxy, as indicated in Table \ref{tab:resonances}. The corotation of the three spiral resonances are located at radii of $\sim$ 50\arcsec, $\sim$ 80\arcsec and $\sim$ 130\arcsec, and the corresponding pattern speeds are $\sim$ 90 $kms^{-1}kpc^{-1}$, $\sim$ 60 $kms^{-1}kpc^{-1}$, and $\sim$ 40 $kms^{-1}kpc^{-1}$, respectively. The ratio of the corotation radii of each pair of two contiguous resonances takes a constant value of 1.6, while the ratio of the pattern speeds takes a fixed value of 1.5. This pattern is also repeated between the bar resonance with a radius of 31\arcsec and a angular rate of 135 $kms^{-1}kpc^{-1}$, and the innermost arm resonance, but not for the three resonances associated with the kinks. We interpret these tight relationships between the four resonances as  evidence that these four describe the basic $m = 2$ resonant structure of the galactic disk. Consequently we infer that the three resonances associated with the kinks have been triggered by the interaction of M51 with its galaxy companion.

\begin{table}[ht]
    \centering
        \caption{Resonances of M51}
    \label{tab:resonances}
    \begin{tabular}{c c c c c}
    \hline \hline
      $r_{CR}(H\alpha)$  & $\Omega_P(H\alpha)$ &  $r_{CR}(CO)$  & $\Omega_P(CO)$  & Associated with \\
      $(arcsec)$ & $(kms^{-1}kpc^{-1})$  & $(arcsec)$ & $(kms^{-1}kpc^{-1})$ & {} \\
      \hline
       - & - & $31.2\pm 3.6$ & $135.0_{-11.6}^{+14.1}$ & Stellar bar\\
       - & - & $49.5\pm 5.3$ & $91.4_{-7.9}^{+9.6}$ & Spiral arms\\
       $62.8\pm 2.8$ & $76.2 \pm 2.1$ & $65.2\pm 3.7$ & $71.4_{-3.6}^{+3.9}$ & weak kink\\ 
       $81.1\pm 4.9$ & $60.7\pm 3.2$ & $77.5\pm 4.9$ & $60.9_{-3.4}^{+3.8}$ & Spiral arms\\ 
    $130.0\pm 4.1$ & $38.7\pm 2.2$ & - & -  & Spiral arms\\ 
       $160.2\pm 1.7$ & $31.4\pm 1.7$ & - & -  & kink Northern arm\\ 
       $188.7\pm 5.6$ & $26.6\pm 1.2$ & - & -  & kink Southern arm\\
       \hline
    \end{tabular}
\end{table}

\begin{table}[ht]
    \centering
        \caption{Comparison of pattern speed values}
    \label{tab:patternspeed}
    \begin{tabular}{c c c}
    \hline \hline
     Reference & Type of data    &  $\Omega_P$ \\
         {} & {}  & $(kms^{-1}kpc^{-1})$\\
         \hline
         \cite{tully_kinematics_1974} &  Simulations &  43.6 \\
         \cite{garcia-burillo_co_1993} & CO & 31\\
         \cite{salo_n-body_2000-1} & Simulations & 58\\
         \cite{zimmer_pattern_2004} & CO & $44\pm 7$\\
         \cite{hernandez_bar_2004}& H$\alpha$ & $46\pm 8$\\
         \cite{egusa_determining_2009} & H$\alpha$ \& CO & $37\pm 5$ \\
         \cite{meidt_radial_2008} & CO &  $104_{-27}^{+20}$, $58_{-11}^{+9}$, 23\\
         \cite{dobbs_simulations_2010}  & Simulations  & between 29 and 58\\
         \cite{querejeta_gravitational_2016} & CO & $168\pm 15, 48\pm 5$\\
         \cite{ghosh_dynamical_2016}  &  Simulations &  between 39.0 and 51.6\\
         \cite{abdeen_determining_2020} & Multi-wavelengths & 21.56 $\pm$ 6.13 \\
         This study & H$\alpha$ & $76.2 \pm 2.1$, $60.7.6\pm 3.2$, $38.7\pm 2.2$, $31.4\pm 1.7$, $26.6\pm 1.2$\\
         {} & CO & $135.0_{-11.6}^{+14.1}$, $91.4_{-7.9}^{+9.6}$, $71.4_{-3.6}^{+3.9}$, $60.9_{-3.4}^{+3.8}$\\
         \hline       
    \end{tabular}
\tablecomments{All values are corrected for a distance of 8.34 Mpc}
\end{table}

\section{Discussion} \label{sec:discussion}

So far, we have been paying full attention uniquely to NGC 5194 (note that in this study we are considering that M51 and NGC 5194 is the same galaxy), which is playing two different roles: the morphology of its spiral arms by means of the pitch angle and the dynamics of the disk by means of the resonant structure. Now, we must welcome a second player, NGC 5195, the companion galaxy of M51, so that the dialogue between these two will provide a fully comprehensive understanding of the entire scene.

\begin{figure}
    \centering
    \plotone{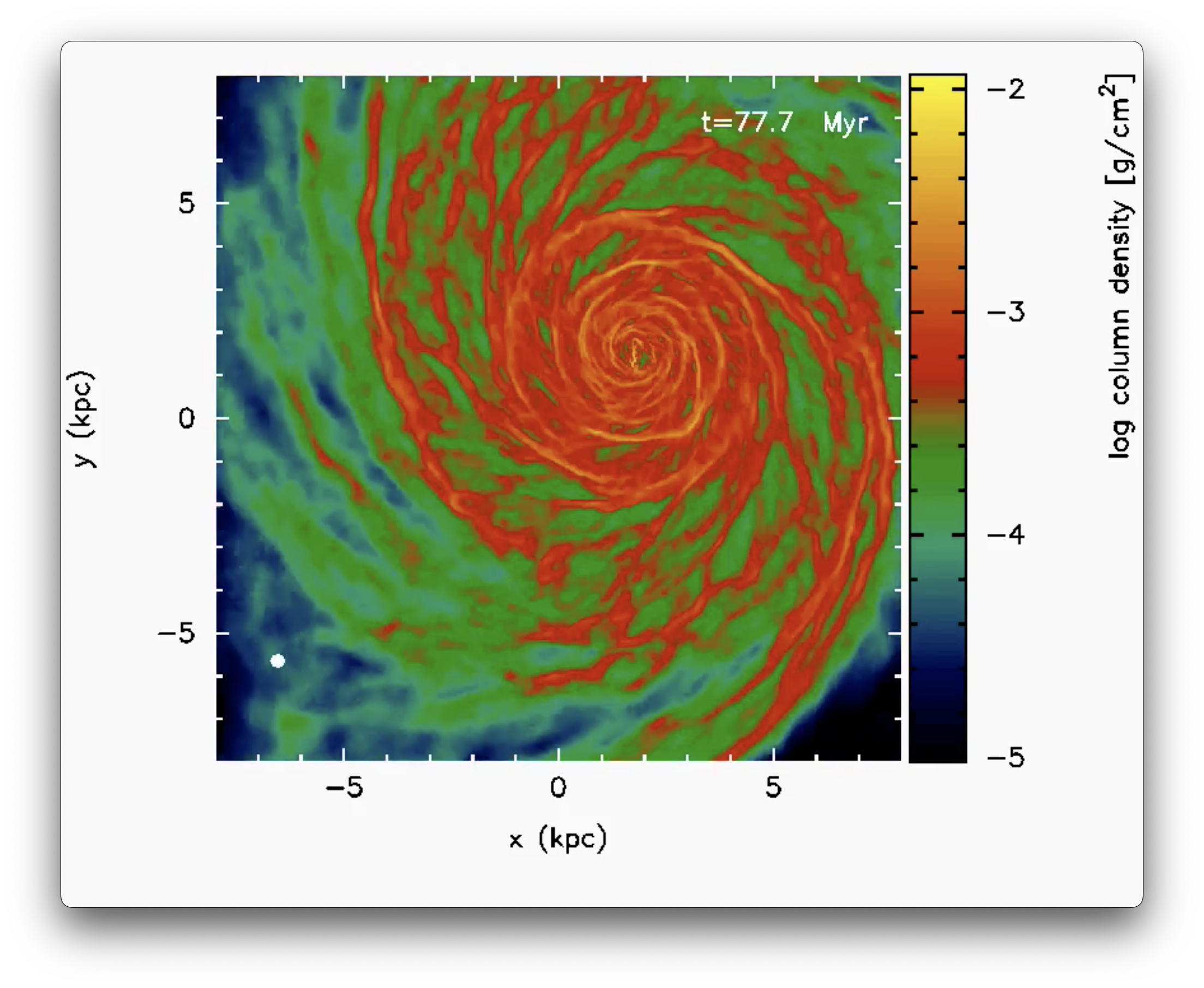}
    \caption{Numerical simulation of the interaction of M51 and NGC 5195, which was also presented in \citep{dobbs_simulations_2010}, where some snapshots were used to compose their Figure 4 (see that paper for a more detailed description of this simulation). The animation, which covers the time scale between $t=0$ My (first frame) and $t=375$ My (last frame) in 9 seconds of real-time video, shows how the gas of a disk galaxy (M51) evolves and forms the spiral arms after the passage of the companion galaxy (shown here as a point mass in white).}
    \label{fig:video}
\end{figure}

In general terms, the formation of the spiral arms can be explained by excitation mechanisms, which can be originated externally or internally. The internal excitation can be produced by the presence of a bar in the center of the galaxy, but also by self-excited instabilities of the disk. The external excitation is generated by the tidal interaction of a companion galaxy (see the reviews by \cite{dobbs_dawes_2014} and \cite{sellwood_spirals_2022} for a detailed description of these mechanisms). M51 is a clear example of a grand design galaxy stimulated by an external perturbation mechanism resulting from the interaction with NGC 5195, and this is supported by numerical simulations of the interaction between M51 and its companion galaxy, showing how M51 has been affected by the passage of the satellite galaxy NGC5195. \cite{toomre_galactic_1972} investigated the effects on the morphology, in terms of tails and bridges, that a galaxy will experience due to different modes of interaction with a companion galaxy, and compare their results to the duo M51-NGC5195. However, the most complete numerical model of M51 and NGC5195 was carried out by \cite{salo_n-body_2000-1,salo_n-body_2000}. These authors demonstrated that for a parabolic single passage or a bound encounter, in which the companion can cross the galactic plane of the primary disk, either type of orbit can reproduce the morphology of M51, whereas only the bound model with multiple encounters can replicate the observed kinematics of M51. These results were confirmed by \cite{theis_m51_2003} who used a different technique to investigate $\sim10^5$ models to derive the optimal region in parameter space. Subsequently, \cite{dobbs_simulations_2010} performed a suite of numerical simulations of M51 in which the gaseous component was included, and they found that with two encounters in which the companion galaxy crosses the galactic plane of M51, the galaxy evolves from a flocculent galaxy to a grand design galaxy in agreement with the morphology of M51 observed, now including the kinks in the spiral arms. This is nicely illustrated in Fig. 4 of \cite{dobbs_simulations_2010}, which shows how the first passage of NGC 5195 gives rise to the spiral arms of M51. This can also be seen in the animation of Fig. \ref{fig:video}. Kinks in both spiral arms are also reproduced in some models of the simulations of \cite{pettitt_bars_2018}.

In this scenario, the first passage of NGC 5195 triggers the two spiral arm structure of M51 \citep[see the second panel of Fig. 4 of][]{dobbs_simulations_2010}. An inference from the simulation of \cite{dobbs_simulations_2010} is that the first encounter affects the central region of M51, and mainly the southern arm (see the second panel of their Fig. 4), and this can explain why the phase reversals associated with the resonances produced in this encounter are mostly located around the southern arm rather the northern one, as illustrated in green and yellow in Fig. \ref{fig:PR2DHa} for the ionized gas only, and in cyan, green and yellow in Fig. \ref{fig:PR2DCO} for the molecular gas only and also in Fig. \ref{fig:PR2DSpitzer} for the two components. Similarly, Fig. \ref{fig:PR2DHa} shows that the two outermost resonances we find for the ionized gas distribute their phase reversals mainly around the northern arm, which is in agreement with a second passage of NGC 5195, as illustrated in the fourth panel of Fig. 4 in \cite{dobbs_simulations_2010}, showing that the companion galaxy excites mainly the external region of the northern arm. This second encounter with NGC 5195 creates a perturbation in the gravitational potential of the outer region of M51, which is responsible for the kink and the subsequent perturbation region that we identify in each spiral arm (see Fig. \ref{fig:N&S}); this perturbation is propagating in each arm at a different angular velocity, therefore a new resonance, which is almost touching the kink, is associated with each spiral arm (as discussed in the previous section). This is consistent with the observational inference that the timescale associated with the perturbation region, $\Delta\tau$ in Myr, that we have estimated using the pattern speed of the resonance and the azimuthal extension of the perturbation region, is the same for both arms.  Furthermore, the change of the measured pitch angle at the kink and at the outer region (see Table \ref{tab:pitchangleM51}) is found to be same in both spiral arms, which also favors this scenario.

\begin{figure}
    \centering
    \includegraphics[scale=1.0]{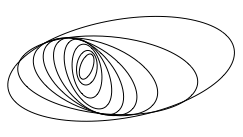}
    \caption{Diagram of elliptical orbits with an asymmetric extension added in order to simulate the case of M51, illustrating how two spiral arms with slightly different pitch angle can be produced, each of them with its own pattern speed and corotation radius.}
    \label{fig:orbits}
\end{figure}

Observational studies of the effects of density waves in the origin and maintenance of spiral arms in disk galaxies have concentrated on deriving a coherent pattern speed, most often associated with a bar, and using the frequency curve, which is obtained dividing the rotation curve (in $(kms^{-1}$) by the radius (in $kpc$), have then derived a corotation radius. A virtue of the FB method is to be able to derive the two- dimensional wave patterns in disks, and this has proved of particular value for M51 where the spiral structure has clearly been stimulated and affected by its interaction with NGC 5195, leading to specific interaction features and to departures from symmetry. Finding two slightly different corotation radii, each associated with one of the spiral arms, can be very well accounted for if the density wave pattern departs from ideal circularity. This can be visualised using a diagram derived from the classical simplified orbital description of density wave structure by Lindblad, in which the individual orbits are no longer simple ellipses, but are distorted due to the interaction with the satellite galaxy; we show this in Fig. \ref{fig:orbits}. A highly plausible rendering of the effects of the interaction can be seen in the animation of Fig. \ref{fig:video}, which was produced by \cite{dobbs_simulations_2010}. And also \citep{2018AJ....156..123A}

\section{Summary} \label{sec:conclusions}

\renewcommand{\labelenumi}{ (\alph{enumi})}

In this study, we have measured and analyzed the morphology of the spiral arms of M51 using the technique of plotting their projection in the log\,($r$)--$\theta$ plane. By using this for each arm, we have shown that:
\begin{enumerate}
    \item M51 is a clear example of a grand design galaxy, as its arms are symmetrical with an azimuthal offset of 180$^{\circ}$.
    \item The pitch angles of the arms are very similar, with the southern arm slightly more tightly wound than the northern arm.
    \item The two arms each show a couple of kinks; while the weak ones are located in the inner region and show a clear symmetry in radius, the main kinks are measured at somewhat different radial and azimuthal positions, indicating that the two kinks in each arm have different mechanism of formation (respond to different processes)
    \item Just beyond each outer kink, the two arms show the same behaviour: they have a perturbation region, which is best described using a quadratic function.
    \item Assuming a linear approximation along the whole length of the spiral arms, the change of the pitch angle at the main kink is very similar in both arms, as is also the change of the pitch angle beyond the kink.
\end{enumerate}

\renewcommand{\labelenumi}{ (\arabic{enumi})}

 Proceeding to use the relevant kinematic information, with very precise velocity maps in H$\alpha$ and CO, we have derived the resonant structure of the Whirlpool galaxy, showing that

\begin{enumerate}
    \item The H$\alpha$ map covers the whole extension of the galaxy but lacks the angular resolution to resolve correctly the central region of the galaxy (r $\lesssim$ 50 \arcsec\ ) as illustrated in  Fig. \ref{fig:PR2DHa}. On the other hand the CO data cover only the central region and do not provide reliable results beyond r $\sim$ 80 \arcsec\ (see figure \ref{fig:PR2DCO}).
    \item When we overplot the phase reversals identified according to the FB method on each corresponding intensity map, we confirm that they are located around the spiral arms.
    \item The H$\alpha$ data reveal five resonances, while the CO map show the presence of four resonances in the central region of the galaxy.
    \item The innermost resonance (in CO) is associated with the bar, giving a rotational parameter (resonance radius divided by bar length) of 1.48.
    \item In the overlapping region between the two data sets, the two innermost resonances in H$\alpha$ are compatible with the two outermost resonances in CO, as their corotation radii are matched. In terms  of pattern speed both of these two resonances show consistent values within the uncertainties, whether measured via the molecular or the ionized atomic gas.
    \item We identify three different resonances associated with the two spiral arms of M51, which define a precise ratio of pattern speeds, as well as corotation radius, between pairs of contiguous resonances, with a value of 1.5 and 1.6, respectively; including the resonance associated with the bar this ratio holds for four contiguous pairs. 
    \item The relative position between the kink and the corotation radius in the log\,($r$)--$\theta$ plane gives a criterion for associating a given resonance with a given kink. Using this criterion, we relate three resonances with the kinks we identify in the spiral arms of M51.
    \item  We find that the timescale of the perturbation region (see Fig. \ref{fig:armfit}) is the same in both arms when using the corresponding pattern speeds for each arm, we interpret this as indicating that the same mechanism is responsible for the production of the kink in each spiral arm.
\end{enumerate}

The third piece of this jigsaw, is the effect of the tidal interaction of the companion galaxy, NGC 5195: 

\begin{itemize}
    \item First encounter: This triggers the formation of the two spiral arms structure, which evolves to a grand design morphology in which the two spiral arms have a very similar pitch angle. The relative orbital position between the two galaxies, according to the numerical simulations, explains the location of the phase reversals we find in M51. This encounter can also be related to the weak perturbation, which we identify in the inner region.
    \item Second encounter: This creates a perturbation, which makes the two spiral arms bend inwards, which we identify as a strong kink, followed by a region (perturbation region) where the spiral arms tend to recover the spiral pattern. From numerical models of the interaction of the two galaxies, we see that the perturbation excites mainly the external region of the northern arm, which is where we find most of the phase reversals.
\end{itemize}

\section{Conclusions}

In this study we have shown how the morphology and the kinematics of a disk galaxy can be combined with the predictions of simulations of that galaxy, to explain quantitatively how the tidal interaction of a companion galaxy has contributed to the formation and structure of its arms. For M51 the measured morphological parameters include the fundamental pitch angles of its two arms, and the morphology and positions of the kinks in the arms caused by two passages of the companion through the plane of the disk. The measured kinematic parameters are the positions in the plane of the disk of the resonances marked by the change from inflow to outflow, or viceversa, of the interstellar gas, both ionized and molecular, in  the spiral arms.

For this grand design galaxy we have shown, combining the observations with simulations, that a single external perturbation of the gravitational potential has created an opposed pair of resonances, and that the presence of another, albeit weaker pair, implies that the companion galaxy has made two encounters with M51.

We have found seven resonances within the disk of M51, one associated with the two weak kinks, and two associated with the strong kinks in the arms,  all of which are the responses to the gravitational perturbations caused by the interaction of NGC 5195. The other four resonances can be associated with the bar: one not far outside the end of the bar, whose radius, at 1.48 times the bar radius, is well within the range found for the bar resonances of many disk galaxies, and three within the spiral arms. The invariant ratios between the radii and the pattern speeds of adjacent pairs of these four resonances leads us to associate the three arm resonances and the bar resonance with an underlying morphological structure throughout the disk due to a perturbation with $m = 2$ symmetry.

These results for M51 are in agreement with the results, both theoretical and observational, of previous studies, showing that disk galaxies can have multiple resonances which are radially distributed, but can also be distributed in azimuth. The spiral arms may have a set of pattern speeds in defined annular zones, with values which fall with increasing galactocentric radius of each annular zone. The spiral arms of galaxies in general have, according to this scenario, both transient and persistent characteristics, because although each annular arm segment may be treated as dynamically separate, there are usually resonant links between the annuli.

The scenario we have found in M51, where the resonances are distributed radially and azimuthally (the two outer resonances are assigned to a different arm), raises questions which require further detailed observational study.

\renewcommand{\labelenumi}{ (\roman{enumi})}

\begin{enumerate}
    \item To what extent does this distribution of resonances occur only in grand design galaxies, or can the pattern be found in multi-arm galaxies?
    \item Is this distribution of resonances induced only by tidal interactions or can it be found in spiral arms formed by internal excitation (bar driven mechanisms, for example)? Results to be found in our previous work suggest that the latter is probably correct. 
    \item Can morphological breaks in spiral arms be produced by internal asymmetries? Here also we venture to suggest that this may indeed be the case.
\end{enumerate}

The global agreement between our observations and the simulations of \cite{dobbs_simulations_2010} should give encouragement to both observers and simulators of galaxy evolution.

\begin{acknowledgments}
This research has made use of the NASA/IPAC Infrared Science Archive and the Extragalactic Database (NED), which are funded by the National Aeronautics and Space Administration and operated by the California Institute of Technology.
\end{acknowledgments}

%


\bibliography{M51_joanfont.bib}{}
\bibliographystyle{aasjournal}



\end{document}